\documentclass[12pt,preprint]{aastex}
        
\def\etal{{\it et~al. }}

\received{2003 August 4}
\begin{document}

\title{A Face--On Tully-Fisher Relation}

\author{David R. Andersen\altaffilmark{1}}
\affil{Max Planck Institute for Astronomy; K\"onigstuhl 17, D-69117
Heidelberg, Germany; andersen@mpia-hd.mpg.de}
\author{Matthew A. Bershady} \affil{Department of Astronomy, University of
Wisconsin, 475 N Charter Street, Madison, WI 53706; mab@astro.wisc.edu}

\altaffiltext{1}{
Visiting Astronomer, Kitt Peak National Observatory, National Optical
Astronomy Observatory, which is operated by the Association of Universities
for Research in Astronomy, Inc. (AURA) under cooperative agreement with the 
National Science Foundation.}

\begin{abstract}

We construct the first ``face-on'' Tully-Fisher (TF) relation for 24
galaxies with inclinations between 16$^\circ$ and 41$^\circ$. The
enabling measurements are integral-field, echelle spectroscopy from
the WIYN 3.5m telescope, which yield accurate kinematic estimates of
disk inclination to $\sim15^\circ$. Kinematic inclinations are of
sufficient accuracy that our measured TF scatter of 0.42 mag is
comparable to other surveys even without internal-absorption
corrections. Three of four galaxies with significant kinematic and
photometric asymmetries also have the largest deviations from our TF
relation, suggesting that asymmetries make an important contribution
to TF scatter. By measuring inclinations below 40$^\circ$, we establish
a direct path to linking this scatter to the unprojected structure of
disks and making non-degenerate dynamical mass-decompositions of
spiral galaxies.

\end{abstract}


\section{Introduction}

One common characteristic of Tully-Fisher (TF; Tully \& Fisher 1977)
studies of the relation between luminosity and rotation speed is the
selection of spiral galaxies with inclinations ($i$) above $45^\circ$,
and typically with $i\sim60^\circ$.  The inability to determine
accurate and precise inclinations for galaxies with $i< 40^\circ$ has
limited studies attempting to link structural parameters, such as
lopsidedness and ellipticity, to TF scatter (Franx \& de Zeeuw 1992,
Zaritsky \& Rix 1997). Likewise, combining vertical stellar velocity
dispersions and disk rotation speeds has been thwarted for lack of
precise inclinations for nearly face-on galaxies (e.g., Bottema
1997). Hence, non-degenerate dynamical mass-decompositions of galaxies
have not yet been made.

One solution to the above problems is to use nearly face-on galaxies
to study the TF relation. Here
the detailed planar structure of the disk can be studied unprojected,
while the vertical component of the disk stellar velocity dispersion
is favorably projected. The problem is now observational: How can
accurate and precise inclinations be measured for small inclination
angles?

For most TF studies, inclinations are calculated from photometric
axis-ratios, assuming circular disks, and correcting for a constant,
intrinsic flattening (e.g. Tully \& Fouqu\'e 1985; Courteau 1996).
Bars, spiral structure, and intrinsic ellipticity cause systematic
errors in photometric inclination and position angles (PA), dominate
the apparent ellipticity at low inclination, and skew apparent
inclinations to larger values (Andersen \etal 2001; Courteau \etal
2003). Hence there is a triple penalty at low inclinations for TF
studies: (1) Erroneously large inclinations lead to under-correcting
the velocity projection. (2) Random errors diverge since photometric
disk axis ratio measurements have constant errors as a function of
axis ratio. (3) For long-slit studies, PA-mismatch will always yield
systematic underestimates of the projected rotation speed.

These pitfalls can be mitigated if kinematic inclinations are derived
from two-dimensional velocity fields (e.g., Schommer \etal 1993;
Verheijen 2001) because disk scale heights are irrelevant, and
velocity fields between 1 and 3 scale lengths are typically less
aberrated than the light distribution by spiral structure, bars or
warps (Courteau \etal 2003; Briggs 1990).
A limit of $\sim40^\circ$ had been set on the minimum derived
kinematic inclinations for nearby galaxies from HI maps (Begeman
1989), but this minimum value is not fundamental; it is notionally
adopted as such.

We have discovered that kinematic inclinations, derived from
high-quality, optical kinematic data collected with integral field
units, are sufficiently accurate and precise to construct a TF
relation at inclinations as low as 15$^\circ$. In this letter we
present an analysis of the first TF relation for a sample of nearly
face-on galaxies.

\section{The Photometric and Kinematic Data Set}

We obtained $R$ and $I$-band images and H$\alpha$ velocity fields of
39 spiral galaxies meeting the criteria of Andersen \etal (2001):
Galaxies appear nearly face-on ($i<30^\circ$), are of intermediate
type (Sab-Sd), with no clear signs of interactions, bars or rings, and
are in regions of low Galactic extinction.  Our sample and
observations are described in Andersen (2001).

CCD images were acquired over 17 nights between May, 1998 and January,
2001 at the WIYN 3.5m\footnote{The WIYN Observatory is a joint
facility of the U. of Wisconsin-Madison, Indiana U., Yale, and NOAO.},
the KPNO 2.1m, and the McDonald Observatory 2.7m telescopes. Total
magnitudes are defined as asymptotic fluxes, as measured from growth
curves, calibrated on the Kron-Cousins system using Landolt (1983)
standards observed on four photometric nights. Zeropoints for other
nights are internally boot-strapped, where possible, from repeat
observations. We use $H_0=70$ km~s$^{-1}$~Mpc$^{-1}$ and correct
magnitudes for Galactic absorption (Schlegel, Finkbeiner \& Davis
1998). (Virgo in-fall affects recession velocities by $<$1\%; Paturel
\etal 1997.) We use the mean $R-I$ color of $0.52\pm0.08$ mag for our
calibrated sample to estimate the $R$ magnitude for several galaxies
with only accurate $I$-band photometry, and add the 0.08 mag range in
$R-I$ in quadrature as an additional uncertainty for these sources.

H$\alpha$ velocity fields were obtained at an instrumental resolution
of ${\lambda\over\Delta\lambda}\approx 13,000$ using the DensePak
integral field unit (Barden, Sawyer \& Honeycutt 1998) on the WIYN
3.5m telescope over 11 nights between May, 1998 and January,
2000. [PGC 56010, a low surface-brightness galaxy, was re-observed
using SparsePak (Bershady \etal 2003) in May, 2001.] The H$\alpha$
emission lines were resolved, with typical signal-to-noise levels of
$\sim35$ and centroid precision of $\sim1.5$ km~s$^{-1}$. DensePak
observations yield velocity centroids for $\sim140$ independent
positions within each galaxy out to 2.2 to 5.8 disk scale lengths
(typically 3.3).

\section{Measuring Inclinations Below $40^\circ$}

The most critical measurement required to construct a ``face-on'' TF
(FOTF) relation is the derivation of accurate and precise disk
inclinations.  As we have argued, there are fundamental reasons why
two-dimensional kinematic maps provide the only viable data for
measuring these inclinations. Using a method similar to that described
in Andersen \etal (2001), we fit the DensePak data with a single,
hyperbolic-tangent velocity-field model. From these fits, we derive
measurements of kinematic inclinations, rotation speeds ($V_{\rm
rot}$), and recession velocities for our sample.

Of the 39 galaxies in our sample we measure reliable deprojected
rotation velocities, $V_{\rm circ}\equiv {V_{\rm rot}/ \sin i}$, for
26 (see Table 1; two galaxies lacking calibrated $R$ and $I$-band
magnitudes are excluded). This subsample has $V_{\rm rot}<120$
km~s$^{-1}$ with inclinations and errors from
16$^{\circ}$$^{+5}_{-10}$ to 41$^{\circ}$$^{+5}_{-6}$. Ten additional
galaxies have measured, deprojected rotation velocities, but kinematic
inclinations consistent with zero within 68\% confidence
intervals. Their mean inclination is $8^\circ$. Three remaining
galaxies have no projected circular velocity field; they appear truly
face-on. Our results paint a consistent picture of being able to
measure inclinations reliably down to $i\sim15^\circ$.

Why can we measure accurate inclinations for nearly face-on galaxies?
For disks with inclinations between $20^\circ$ and $30^\circ$, the
peak difference in projected velocity is only 2 to 3 km~s$^{-1}$
between disks with inclinations differing by 5$^\circ$. To measure
this difference requires $<2$ km~s$^{-1}$ centroiding precision per
fiber, or many samples of the velocity field at a precision at least
comparable to this difference. The latter, which we achieve, is
preferable because of random motions. We find typically 3 km~s$^{-1}$
rms variations between our model and the data (Andersen 2001).

Historically, HI data have been claimed to be of insufficient quality
for deriving galaxy inclinations below 40$^\circ$ (Begeman 1989). One
aspect of the problem has been that tilted ring models include a
weighting scheme which, when typically implemented, minimizes the
importance of velocity measurements at azimuthal angles which are most
influenced by changes in inclination. For example, Begeman's
simulations use a $\cos \theta$ weighting scheme, where $\theta$ is
the angle from the major axis in the galaxy plane. This weighting is
done to minimize the effects of warps, but in so doing over half of
the signal is thrown out. For the same reason, data are often not
considered beyond $\theta$ of 30$^\circ$ or 45$^\circ$, throwing out
yet more of the signal.

Our DensePak observations sample the inner disks of galaxies, where
warps tend to be negligible. Hence we can use the entire velocity field to
maximize the influence of inclination in the model velocity
fields. Figure 1 shows deprojected position-velocity diagrams for two
representative galaxies in  different bins of $\theta$. Excellent
agreement between azimuthal bins indicates a good model fit and lack
of systematics. We find bidimensional echelle spectroscopy to be of
sufficient quality to yield deprojected rotation velocities, $V_{\rm
rot}/ \sin i$ at precisions better than 10\% at inclinations above
28$^\circ$; a 20\% precision on $V_{\rm rot}/ \sin i$ is reached for
inclinations of $\sim20^\circ$.

\section{FOTF vs Conventional TF Relations}

To evaluate our FOTF data, we compare to the TF samples of Courteau
(1997) and Verheijen (2001).  The salient difference between our
sample versus Courteau's and Verheijen's is that their galaxies have
$i>40^\circ$ and $\overline{i} \sim65^\circ$. Courteau's sample shares
comparable scale lengths, colors, surface brightnesses, distances and
uses spatially-resolved H$\alpha$ long-slit emission as the kinematic
tracer.  We use the cleaner ``Quiet Hubble Flow'' (QHF) subset of
Courteau's sample and convert Courteau's Lick $r$-band photometry to
Cousins $R$-band using $R-r=-0.36$ (Fukugita \etal 1995). Verheijen
collected multi-band (including Cousins $R$-band) photometry and HI
synthesis maps for a volume-limited sample from the nearby Ursa Major
cluster. We select the Ursa Major subsample with $M_R
< -19$, corresponding to the faint end of the FOTF and Courteau
samples, and adopt the HST Key Project distance (Sakai \etal 2000).
Both Courteau and Verheijen use an internal-absorption correction of
the form $A_i = \gamma \log a/b$ which minimizes the apparent TF
scatter. (We have subtracted from Courteau's data his zeropoint
correction of 0.40 mag, which corresponds to his absorption correction
at $i=70^\circ$.)  Since our sample is nearly face-on, we apply no
internal-absorption corrections to our photometry.

An inspection of the three samples in $M_R$ versus $\log(2 V_{rot}/
\sin i)$ of Figure 2 highlights the differences and similarities in
their TF relations. There is excellent agreement between the Courteau
and Verheijen TF slopes and zeropoints. There is also an excellent
agreement in the zeropoint of the FOTF with Verheijen and Courteau TF
relations near L* at $M_R\sim-22$, but the shallower slope of the FOTF
appears striking. However, slopes vary between $-7$ and $-9$ for both
the Courteau and Verheijen samples depending on the method of velocity
measurement, the extinction correction prescription and the sample
selection. Verheijen (2001) notes the above absorption corrections
steepen TF slopes. However, these corrections do not alter the slope
of the FOTF relation. While empirical corrections (e.g., Giovanelli
\etal 1994, Tully \etal 1998) work well for galaxies with
$40^\circ<i<80^\circ$, the corrections for galaxies with $i<30^\circ$
are small ($A_R\sim0.1$ mag).  Since in practice one can systematize
velocity measurements, a larger FOTF spanning a broader range in
luminosity may provide a new test of the veracity of TF-based
internal-extinction corrections.

    When we restrict our sample by luminosity or $V_{\rm rot}$, we find
    slopes between -5 and -8; a FOTF subsample with $2.3<\log(2V_{\rm circ})
    < 2.7$ shares the same slopes and zeropoints as the Courteau and
    Verheijen culled samples (Figure 1). This implies that only the extreme
    rotators, fast and slow, differ. This may be due to luminosity-dependent
    extinction corrections discussed above. It is more difficult to present
    a plausible systematic error whereby the high luminosity objects rotate
    faster and the low luminosity objects rotate slower, and the intermediate
    objects are unaffected. However, with only six galaxies in these regimes,
    the discrepancy between slopes may be influenced by our small-number
    statistics coupled with relatively large errors on $\log V/\sin i$.

  If the FOTF is to be a useful tool for studying internal extinction
  or, e.g., correlations of galaxy properties to TF scatter, the impact
  of inclination measurements on TF residuals needs to be
  examined. Figure 3 shows the relation between TF residuals and
  inclination along with the predicted envelope of TF residuals
  introduced by inclination error. We find the scatter for our sample is
  comparable to the scatter in the culled Courteau and Verheijen
  samples: 0.42 mag for ours versus 0.38 mag for the Verheijen and
  Courteau samples used here, each about their respective regressions
  (Figure 2). Six galaxies labeled in Figures 2 and 3 are noted in Table
  1 as either being photometrically or kinematically asymmetric, or
  having small recession velocities (hence uncertain
  distances). Inclination errors alone aren't the cause of their large
  TF residuals. Removing these six galaxies, the scatter for our face-on
  sample about its shallow regression drops significantly to 0.3 mag,
  and is consistent with our measurement errors.  

\section{SUMMARY AND DISCUSSION}

We have presented a TF relationship for 24 galaxies with a mean
inclination of 26$^\circ$ -- the first time the TF relation has been
measured for a sample of galaxies with inclinations less than
$40^\circ$. Accurate and precise kinematic inclinations are the key to
studying TF relation at low inclination, and velocity fields are the
essential ingredient for their measurement.  We are able to construct
this FOTF relation from integral-field H$\alpha$ velocity fields for
four primary reasons: (1) DensePak and SparsePak are efficient for
gathering H$\alpha$ velocity fields with moderately high spatial
resolution and medium spectral resolution at high signal to noise. (2)
H$\alpha$ velocity fields generally extend out to $\sim3$ disk
scale-lengths so that warping, which dominates the outer radii of some
galaxy disks, is unimportant. (3) Instead of using multiple tilted
rings, we successfully fit galaxy disks with a single velocity-field
model, simultaneously using data at all radii and phase angles. (4) We
equally-weight all data, making our fits sensitive to the signal
constraining inclination, which peaks in the velocity field at
$\sim45^\circ$ from the major axis.

The FOTF relation measured from our data has a zeropoint and scatter
comparable to conventional TF relations but possibly a shallower slope. The
bulk of the FOTF scatter is consistent with measurement errors, as
found by Verheijen (2001). Of the four galaxies which show large
kinematic and photometric asymmetries, three fall well off the TF
relation. While our statistics are small, this is consistent with the
claim by Barton \etal (2000) that kinematic asymmetries are an
important source of TF scatter.

The FOTF can play an important role in understanding the nature and
interpretation of the TF slope, zeropoint and scatter in three ways.
First, internal extinction is less important for face-on galaxies
minimizing inclination-dependent effects.  For inclined TF samples, if
correlated errors exist between de-reddened luminosities and
de-projected rotation-speeds, TF scatter will be increased. If
extinction depends on luminosity or velocity (mass), the corrected TF
slopes may differ from their true values.

Second, measurements of asymmetry, ellipticity, and lopsidedness
ideally are made at low inclinations. Several studies have suggested
non-axisymmetric dark halo mass distributions such as lopsidedness
(Zaritsky \& Rix 1997; Swaters \etal 1999) 
or ellipticity (Franx \& de Zeeuw 1992;
Andersen \etal 2001) could be sources of TF scatter. However, without
access to inclination measurements for face-on systems, these claims
could not be directly evaluated. Our small FOTF sample suggests such
links are possible, and show that such studies can now be pursued
wholesale.

Finally, the mass-to-light ratio of spiral disks is an essential
ingredient for interpreting the TF zeropoint, slope, and scatter. Only
limits can be placed on mass decompositions (Bell \& de Jong 2001)
without direct dynamical estimates of total {\it and} disk mass. This
can be accomplished via rotation curves {\it and} vertical stellar
velocity dispersions of disks. The latter are best measured in face-on
systems, and hence the FOTF unlocks the door for determining the
fundamental mass-budget of spiral galaxies.

\acknowledgments

We thank E. Bell, S. Courteau, and B. Tully for helpful
comments.
We acknowledge NSF AST-9970780 and AST-0307417.

\bibliographystyle{apj}

\begin{thebibliography}{27}
\expandafter\ifx\csname natexlab\endcsname\relax\def\natexlab#1{#1}\fi

\bibitem[]{357} Akritas, M.G. \& Bershady, M.A. 1996, \apj, 470, 706

\bibitem[]{359} Andersen, D. R. 2001, Ph.D. thesis, Penn State University

\bibitem[]{andersen01} Andersen, D.R., Bershady, M.A., Sparke, L.S.,
Gallagher, J.S. \& Wilcots, E.M. 2001, \apjl, 551, 131

\bibitem[{{Barden} {et~al.}(1998){Barden}, {Sawyer}, \& {Honeycutt}}]{barden98}
{Barden}, S.~C., {Sawyer}, D.~G., \& {Honeycutt}, R.~K. 1998, SPIE, 3355, 892

\bibitem[]{367} Barton, E.J., Geller, M.J., Bromley, B.C., van Zee, L., 
Kenyon, S.J. 2001, \aj, 121, 625

\bibitem[]{begeman89} Begeman, K. 1989, A\&A, 223, 47

\bibitem[]{346} Bell, E.F. \& de Jong, R.S. 2001, \apj, 550, 212

\bibitem[]{372} Bershady, M.A., Andersen, D.R., Verheijen, M.A.W.,
W, K.B., Crawford, S.M., \& Swaters, R.A. 2003, ApJS submitted

\bibitem[]{375} Bottema, R. 1997, A\& A, 328, 517

\bibitem[]{353} Briggs, F.H. 1990, \apj, 352, 15

\bibitem[]{courteau96} Courteau, S. 1996, \apjs, 103, 363

\bibitem[]{courteau97} Courteau, S. 1997, \aj, 114, 2402

\bibitem[]{381} Courteau, S., Andersen, D.R., Bershady, M.A., MacArthur, L.A.,
Rix, H.-W. 2003, \apj, 594

\bibitem[{{Franx} \& {de Zeeuw}(1992)}]{franx92}
{Franx}, M. \& {De Zeeuw}, T. 1992, \apjl, 392, L47

\bibitem[]{fukugita95} Fukugita, M., Shimasaku, K. \& Ichikawa, T. 1995,
\pasp, 107, 945

\bibitem[]{giovanelli94} Giovanelli, R., Haynes, M.P., Salzer, J.J,
Wegner, G., da Costa, L.N., Freudling, W. 1994, \aj, 107, 2036

\bibitem[]{landold83} Landolt, A.U. 1983, \aj, 88, 853


\bibitem[]{397} Paturel, G. \etal 1997, A\&AS, 124, 109

\bibitem[]{377} Sakai, S., \etal 2000, \apj, 529, 698

\bibitem[]{schlegel98} Schlegel, D.J., Finkbeiner, D.P., Davis, M. 1998,
\apj, 500, 525

\bibitem[]{schommer93} Schommer, R.A., Bothun, G.D., Williams, T.B. \&
Mould, J.R. 1993, \aj, 105, 97

\bibitem[]{385}Swaters, R.A., Schoenmakers, R.H.M., Sancisi, R., van Albada, T.S.
1999, \mnras, 304, 330 

\bibitem[{{Tully} \& {Fisher}(1977)}]{tully77}
{Tully}, R.B. \& {Fisher}, J.R. 1977, A\&A, 54, 661

\bibitem[]{tully85} {Tully}, R.B. \& Fouqu\'e, P. 1985, \apjs, 58, 67

\bibitem[]{393}Tully, R.B., Pierce, M.J., Huang, J.-S., Saunders, W., Verheijen,
M.A.W., Witchalls, P.L. 1998, \aj, 115, 2264

\bibitem[]{verheijen0} Verheijen, M.A.W. 2001, \apj, 563, 694

\bibitem[Zaritsky \& Rix(1997)]{zaritsky97}
{Zaritsky}, D. \& {Rix}, H.-W. 1997, ApJ, 477, 118

\end{thebibliography}

\clearpage

\begin{deluxetable}{rlll}
\tablenum{1}
\tablewidth{0pt}
\tablecaption{Tully-Fisher Parameters}
\tabletail{\hline}
\tablehead{
\colhead{} &
\colhead{$i$} &
\colhead{$V_{\rm circ}$} &
\colhead{$M_R$} \nl
\colhead{PGC} &
\colhead{(deg)} &
\colhead{(km~s$^{-1}$)} &
\colhead{(mag)}
}
\startdata
 2162$^{\;\;}$ & 25.8$_{-6.6}^{+5.2}$ & 81.2$^{+38.5}_{-30.4}$   & -20.42$\pm0.03$ \\
 3512$^{\;\;}$ & 30.1$_{-2.7}^{+2.3}$ & 164.3$^{+26.7}_{-22.8}$   & -21.17$\pm0.03$ \\
 5673$^{\;\;}$ & 21.0$_{-5.0}^{+3.2}$ & 160.6$^{+72.6}_{-46.5}$   & -20.89$\pm0.05$ \\
 6855$^{a}$    & 38.8$_{-3.8}^{+3.2}$ & 84.3$^{+14.0}_{-11.8}$   & -20.64$\pm0.05$ \\
 7826$^{b}$    & 40.7$_{-3.8}^{+3.2}$ &$\;\;$49.0$^{+11.2}_{-8.3}$& -19.30$\pm0.06$ \\
14564$^{a}$    & 30.5$_{-3.1}^{+2.9}$ & 181.8$^{+33.4}_{-31.2}$   & -21.48$\pm0.05$ \\
16274$^{\;\;}$ & 30.6$_{-2.7}^{+2.3}$ & 222.4$^{+35.4}_{-30.2}$   & -22.03$\pm0.05$ \\
23333$^{\;\;}$ & 26.9$_{-4.5}^{+3.8}$ & 133.0$^{+40.9}_{-34.6}$   & -20.67$\pm0.06$ \\
23598$^{\;\;}$ & 16.2$_{-9.8}^{+5.2}$ & 162.3$^{+189.8}_{-100.7}$ & -21.69$\pm0.05$ \\
24788$^{a}$    & 41.5$_{-6.2}^{+4.9}$ & 112.8$^{+27.4}_{-21.7}$   & -21.70$\pm0.05$ \\
26140$^{\;\;}$ & 28.2$_{-4.3}^{+3.0}$ & 317.3$^{+88.1}_{-61.6}$   & -22.86$\pm0.05$ \\
28310$^{\;\;}$ & 35.3$_{-3.2}^{+2.8}$ & 110.9$^{+17.5}_{-15.3}$   & -20.38$\pm0.07$ \\
31159$^{a}$    & 22.6$_{-4.0}^{+3.0}$ & 220.1$^{+73.4}_{-55.0}$   & -21.19$\pm0.07$ \\
32638$^{\;\;}$ & 22.0$_{-4.9}^{+3.7}$ & 234.25$^{+98.4}_{-74.4}$   & -21.80$\pm0.05$ \\
33465$^{\;\;}$ & 18.6$_{-5.6}^{+4.1}$ & 309.8$^{+178.6}_{-130.8}$ & -22.41$\pm0.05$ \\
36925$^{\;\;}$ & 22.6$_{-4.5}^{+3.5}$ & 195.2$^{+73.2}_{-57.0}$   & -22.06$\pm0.07$ \\
38908$^{\;\;}$ & 26.0$_{-4.0}^{+3.6}$ & 211.0$^{+60.3}_{-54.3}$   & -21.96$\pm0.06$ \\
39728$^{b}$    & 31.8$_{-3.6}^{+3.3}$ & 144.2$^{+29.2}_{-26.8}$   & -20.32$\pm0.06$ \\
46767$^{\;\;}$ & 22.7$_{-3.0}^{+2.6}$ & 288.2$^{+43.4}_{-50.6}$   & -22.79$\pm0.07$ \\
56010$^{\;\;}$ & 26.7$_{-13.6}^{+8.3}$& 75.3$^{+70.7}_{-43.2}$   & -19.97$\pm0.07$ \\
57931$^{\;\;}$ & 21.5$_{-5.3}^{+4.0}$ & 192.4$^{+90.0}_{-68.0}$   & -21.83$\pm0.05$ \\
58410$^{\;\;}$ & 29.0$_{-2.5}^{+2.0}$ & 231.4$^{+36.3}_{-29.1}$   & -22.11$\pm0.10$ \\
72144$^{\;\;}$ & 27.9$_{-5.9}^{+4.2}$ & 110.6$^{+42.8}_{-30.6}$   & -21.27$\pm0.03$ \\
72453$^{\;\;}$ & 21.2$_{-6.6}^{+4.0}$ & 212.4$^{+125.6}_{-76.3}$  & -22.23$\pm0.03$ \\
\enddata
\tablenotetext{a}{Significant photometric and kinematic asymmetry.}
\tablenotetext{b}{$V_{Heliocentric} < 3000$ km~s$^{-1}$.} 
\end{deluxetable}

\clearpage

\begin{figure}
\epsscale{0.8}
\plotone{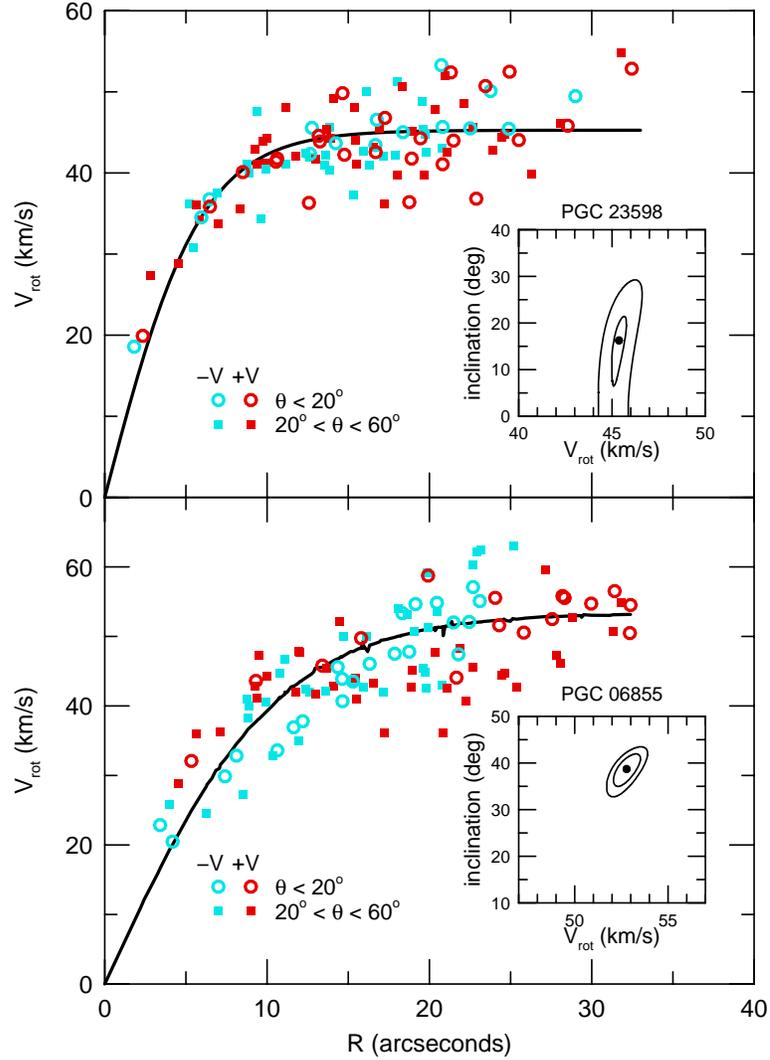}
\caption{Azimuthally-deprojected position-velocity diagram for PGC
23598 (Top) and PGC 6855 (Bottom). Each datum is a separate fiber
measurement. Velocities, color coded as receding (red) or approaching
(blue), are marked as open circles for $\theta < 20^\circ$ and
filled squares for $20^\circ < \theta < 60^\circ$. Solid curves trace
the projected {\it model} rotation curve along the major
axis. $\chi^2$ maps of the 68\% and 90\% confidence limits on model
inclination and rotation velocity are inset. PGC 6855 is
photometrically lopsided and kinematically disturbed; it falls well
off the TF relation (Figure 2). PGC 23598 has the lowest sample
inclination but falls directly on the TF relation.}
\end{figure}

\pagebreak

\begin{figure}
\epsscale{0.7}
\plotone{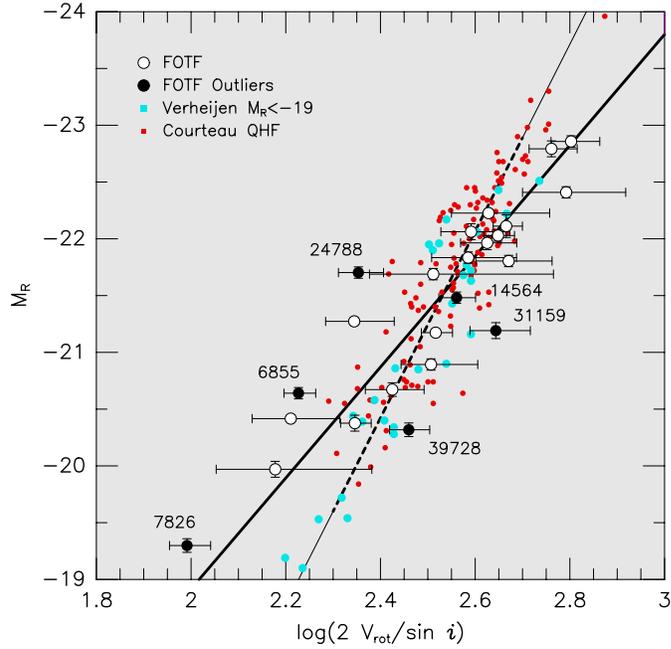} 
\caption{Tully-Fisher relations for our sample of galaxies with 
$\overline{i} = 26^\circ$ (FOTF, white and black) and more-inclined
samples of Courteau (1997; red) and Verheijen (2001; blue). 
FOTF ``outliers'' (filled black circles) are defined in text and
Table~1. The heavy solid line with slope -4.90 is the best fit to the
FOTF data (with or without outliers) using a backward regression which
allows for intrinsic scatter and heteroscedastic errors (BCES[X$|$Y];
Akritas \& Bershady 1996). The heavy dashed line is the best
fit to the FOTF sample between $2.3<\log V_{\rm rot}<2.7$; it is
indistinguishable from the best fits to Courteau and Verheijen (light
solid line with slope -8.2).}
\end{figure}

\pagebreak

\begin{figure}
\epsscale{0.8}
\plotone{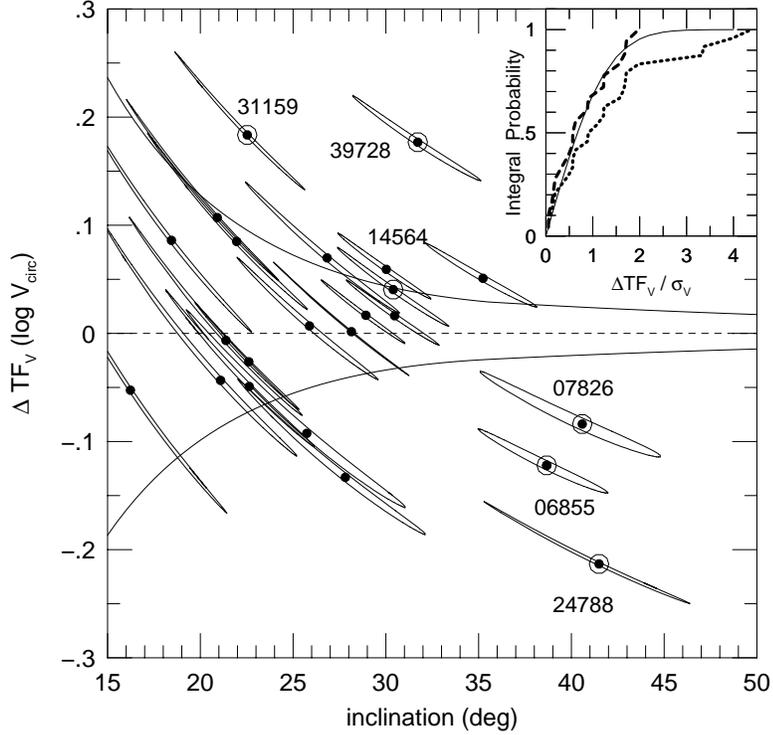}
\caption{TF residuals versus inclination for sample galaxies. 68\%
confidence limits on each galaxy reflect uncertainty in kinematic
inclination and rotation velocity propagated through the best-fit TF
relation. The envelope defined by the two, solid curves enclose the
area in which 68\% of points are expected to fall if {\it all} TF
scatter were due solely to uncertainties in kinematic
inclinations. ``Outlier'' galaxies (see text and Table 1) are
labeled. That 5 of the 6 labeled galaxies fall outside the envelope
suggests that factors besides inclination errors make these galaxies
TF outliers.}
\end{figure}

\end{document}